# Measurements of microturbulence of Cepheids using the autocorrelation function

E.F. Borra & D. Deschatelets *

*Département de Physique, Université Laval, Québec, Qc, Canada G1V 0A6*

**ABSTRACT**

The autocorrelation function (ACF) is a simple and useful tool that allows to measure the average linewidth of spectra with great precision. Unlike the cross-correlation function (CCF), the ACF can be used without having to rely on weighted binary masks. It makes the ACF much easier to use. We analyze 7 Cepheids and present a new method to obtain very precise turbulent velocity curves for 6 of them by using the ACF. We compare our FWHM curves as a function of the pulsational phase to those of Nardetto et al. (2006) who used the Fe I 6056.005 Å line only. We note a significant improvement in the shape of the FWHM curves for all Cepheids by using the ACF. From the FWHM curves, we measure microturbulence curves for 6 Cepheids by using a Gaussian approximation. Finally, we artificially degrade the resolution of the spectra and add noise to further assess and highlight the advantages of the ACF. The FWHM curves obtained with the ACF remain virtually unchanged up to a degradation by a factor of 10. A degradation by a factor of 20 slightly affected the results but an average linewidth variation remains easily detectable.

**Key words:** stars: variables: Cepheids – Turbulence – Techniques: spectroscopic

## 1 INTRODUCTION

Cepheids have been studied and analyzed for quite some time. They are variable stars characterized by a pulsation period generally between 1 and 50 days among the Cepheids. During the pulsation of the star, one can observe a cyclic phenomenon of radial pulsation. The Cepheids contract and expand periodically, thereby generating a variation of the luminosity and the line profiles of the star.

It is well known that Cepheids undergo a phenomenon of line broadening during a pulsation cycle. This linewidth variation has been detected for many Classical Cepheids (e.g. $\delta$ Cephei). The variation of temperature that occurs in Cepheids during the pulsation period does not by itself explain the amplitude of the linewidth variation. In many cases, the ratio between the maximum width of the lines and the minimum width is greater than 50%. If we compare the observed linewidth with a theoretical model taking into account only the pulsational broadening of the Cepheids, we note that there is a a residual broadening that separates the two sets of values.

Some authors have proposed as an explanation that the residual broadening is caused by a movement of turbulence whose intensity varies during the pulsation period of the star (Bersier & Burki 1996; Benz & Mayor 1982; Kovtyukh et al. 2003; van Paradijs 1971). The amplitude of the turbulent velocity is maximum at the phase where the Cepheid is near its minimum radius.

The autocorrelation technique was used by us in a previous article (Borra & Deschatelets 2015) where we conducted a study on magnetic stars. We were able to measure very precise magnetic variation curves as a function of the rotation period of the stars. The autocorrelation function (ACF) also lends itself well to the study of line profile variation of Cepheids. This paper is divided into three main sections. In section 2, we describe the ACF as well as the methodology used to obtain precise turbulent velocity curves from spectra. In section 3, we present the results obtained using the ACF. In section 4, we discuss the results and the conclusions.

## 2 AUTOCORRELATION OF THE SPECTRUM

Our spectra are originally obtained in intensity $I(\lambda)$ as a function of wavelength units (Å). We convert them in velocity $I(v)$ units because we want to use the autocorrelation to obtain microturbulence curves in velocity units (km s$^{-1}$). The conversion is done by interpolating the intensity values over a new series of values with fixed velocity increment per pixel. This new series of values is generated according to Eq. (1):

$$\lambda(i+1) = \lambda(i) + \lambda(i)\frac{v}{c} \tag{1}$$

where $\lambda(i)$ is the wavelength at the $i^{\text{th}}$ pixel, v is the fixed velocity value per pixel and $c$ is the speed of light.

The starting value at position $\lambda(1)$ is 3781.91 Å and the velocity increment is set at 50 m s$^{-1}$. Lower increment values did not provide any improvements in terms of precision and greatly increased the computational time. In all the plots of microturbulence as a function of phase, we use microturbulence in velocity units (km s$^{-1}$). The

* E-mail: borra@phy.ulaval.ca

autocorrelation of the intensity $I(v)$ as a function of the velocity v of the spectra is given by

$$I \otimes I = \int_{-\infty}^{\infty} I(v+v')I(v)dv, \qquad (2)$$

This allows us to calculate an average line profile with a strong signal-to-noise ratio (SNR). The ACF operates in a manner similar to the cross-correlation function (CCF). The advantages of the ACF over the CCF arise from its simplicity and ease of use. The CCF requires the use of a stellar template in which slits are positioned where the lines are present. These slits are weighted according to the depth of the lines. Constructing a weighted binary mask that matches precisely a stellar spectrum is a complicated task. Unlike the cross-correlation function (CCF), the autocorrelation function is directly applicable to spectra using Eq. (2) without using a weighted binary mask. This makes it much easier to use. As described in Borra & Deschatelets (2015), the ACF is particularly efficient against photon noise making it a powerful tool that is well suited to determine with great precision the full width at half maximum (FWHM) of the average single-line profile of the spectrum (see Fig. 1 of Borra & Deschatelets (2015)).

The application of the ACF to the spectra is straightforward. Only a few minor changes need to be made to the spectra. Prior to using the ACF, it is necessary to remove some problematic regions in the spectra that are likely to render the ACF inaccurate because they contain strong lines (e.g. H lines, telluric lines). This can easily be done by setting the intensity to 0.0 in wavelength intervals where such lines are present. The contribution of these lines to the autocorrelation is thereby totally eliminated. In this paper, we model the average line profile of ACF as well as other broadening mechanisms by a Gaussian function. The autocorrelation function is symmetrical and so will be the average line profile from which the FWHM is calculated. Note that this symmetry is generated by the integral in Eq. (2), so that even asymmetrical line profiles (e.g. generated by pulsation) would give a symmetrical autocorrelation profile. The Gaussian function is a very good approximation.

## 2.1 Microturbulence curves

To obtain the microturbulence curve of Cepheids, we use a similar approach to the one used by Bersier & Burki (1996). The observed width of the ACF can be written as follows:

$$\sigma_{obs}^2 = \sigma_{inst}^2 + \sigma_{puls}^2 + \sigma_{res}^2, \qquad (3)$$

where $\sigma$ is the standard deviation of the different broadening mechanisms modeled by a Gaussian function.

We use the same notations as those in Bersier & Burki (1996). $\sigma_{obs}$ is the observed width of the ACF, $\sigma_{inst}$ is the instrumental broadening determined by the resolving power of the spectrograph used, $\sigma_{puls}$ is the broadening due to the pulsational velocity of the Cepheid (expansion/contraction) and $\sigma_{res}$ is the residual width characterized by the rest of the broadening mechanisms such as rotation and microturbulence. The line broadenings caused by microturbulence and rotational velocity have the same wavelength dependence and therefore cannot be identified using the wavelength dependence alone. On the other hand, the time variations of the microturbulence cause the time variations of the linewidth. This is discussed at length in Bersier & Burki (1996) where section 3 discusses the constraints on the rotational velocities. To begin with, let us note that section 3 in Bersier & Burki (1996) concludes that the rotational velocities of the Cepheids are small and that, at the phase of maximum radius, the broadening due to turbulence is the dominant factor. Section 4 in Bersier & Burki (1996) discusses in details the broadening due to turbulence and how it can be estimated from the residual width in Eq. (3). The same techniques can be used to estimate the broadening due to turbulence given by the autocorrelation.

We decide to neglect the instrumental broadening considering the very high spectral resolution of the spectrograph used to obtain the spectra. This particular broadening mechanism will become an important factor as the spectral resolution decreases (see section 3.3).

The scope of this paper is first and foremost to demonstrate the precision of the $\sigma_{obs}$ curve from which $\sigma_{res}$ is calculated using Eq. (3). Broadening due to $\sigma_{puls}$ can be determined by modelization. Bersier & Burki (1996) used a Doppler broadening modeling program to calculate this mechanism as a function of phase.

In our case, we decide to represent the pulsational broadening by a Gaussian function where $\sigma_{puls}$ increases linearly with the value of the radial velocity due to pulsation in the stellar rest frame.

$$\sigma_{puls} = A \cdot |RV_p|, \qquad (4)$$

where A is a weighting factor to be determined and $RV_p$ is the radial velocity in the stellar rest frame of the Cepheid due to its pulsational motion in absolute value as measured by the observer.

In this work, radial velocity values used to calculate $\sigma_{puls}$ are taken directly from Nardetto et al. (2006) who used a bi-gaussian fit to obtain them. These values are represented by the parameter $RV_m$. It is the effective radial velocity of the Cepheids as measured by the observer. To calculate the parameter $RV_p$ in Eq. (4), we need to have the systemic velocity of the Cepheids ($RV_s$) (also known as the center of mass velocity)

$$RV_p = RV_m - RV_s. \qquad (5)$$

Systemic velocity values of the Cepheids ($RV_s$) were taken from the SIMBAD Astronomical Database. These values are very close to those of Nardetto et al. (2006) who used values of the Galactic Cepheid Database (GCD) (see their Table 2). Note that GCD values could also be used but would not lead to any significant changes in our results. We compare the pulsational broadening curves as a function of phase obtained using Eq. (4) with those of Bersier & Burki (1996) who used a modeling program. We vary the weighting factor A between 0 and 1 and select the value that best matches the curves obtained by Bersier & Burki (1996). We find that A = 0.27 is the best matching value. Our discussion in section 3.1 that compares our results to those of Nardetto et al. (2006) validates our use of Eq. (4) and A = 0.27.

## 3 RESULTS

In this section, we first validate the precision of the ACF by comparing our FWHM curves obtained for 6 Cepheids using the autocorrelation function to those of Nardetto et al. (2006) who used the Fe I 6056.005 Å line only (see Fig. 1). We also compare the $\sigma_{obs}$ curve for the Cepheid X Cyg obtained with the ACF to the curve of Bersier & Burki (1996) who used the CCF technique (see Fig. 2). We then present the results ($\sigma_{obs}$ and $\sigma_{res}$ curves) obtained for 6 other Cepheids using the autocorrelation function (see Fig. 3).

### 3.1 FWHM curves of Cepheids

The 6 Cepheids analyzed in this section have a visible magnitude $m_v$ ranging from 5.74 to 7.08. The spectra were obtained with the HARPS spectrograph (R ∼ 120000) by Nardetto et al. (2006)

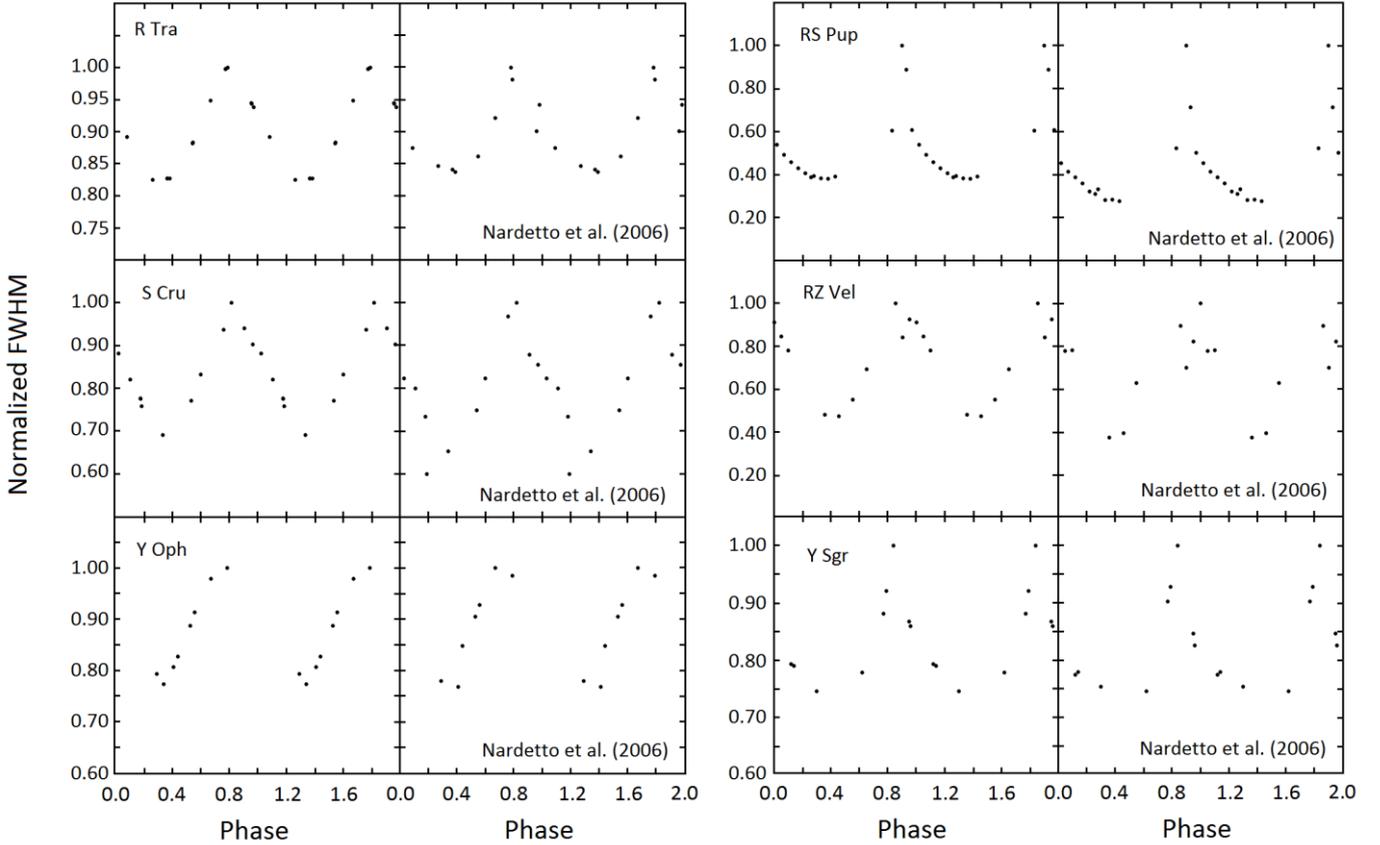

**Figure 1.** Normalized FWHM of 6 Cepheids. Comparison between curves obtained using the autocorrelation function (left side) and curves obtained by Nardetto et al. (2006) using the Fe I 6056.005 Å line only (right side).

and were downloaded directly from the ESO database. Nardetto et al. (2006) analyzed 9 Cepheids in total. We decide to publish results for the 6 fainter Cepheids where statistical uncertainties are the highest. This allow us to better demonstrate the advantage of the ACF over the single-line measurement technique. Note that the 3 other Cepheids ($\beta$ *Dor*, $\zeta$ *Gem* and *lCar*) were also analyzed with the ACF but due to their high apparent luminosity (they are nearby objects), no significant improvement could be observed when compared to the results obtained by Nardetto et al. (2006). The main purpose of this work is to demonstrate the advantage of using the autocorrelation function with noisy spectra.

In Fig. 1, we compare our FWHM curves as a function of phase (left) of the 6 Cepheids analyzed to those of Nardetto et al. (2006) (right). All FWHM curves were normalized to 1. For most Cepheids, the amplitude of the linewidth variation is similar. For the case of R Tra, we measure an amplitude slightly higher than Nardetto et al. (2006). On the other hand, we note a smaller amplitude variation for RS Pup and RZ Vel.

In all cases, we observe a variation pattern that is in agreement with those in Nardetto et al. (2006). However, one can see a considerable improvement in the shape of the FWHM curves for all Cepheids by using the ACF instead of a single line. The curves are less noisy. Most noticeable cases are R Tra, S Cru and RZ Vel. Let us recall that we used the same spectra as those of Nardetto et al. (2006). The precision of the variation curves obtained with the ACF allows to extract accurate measurements of microturbulence velocity as a function of phase.

Statistical uncertainties of the $\sigma_{obs}$ and FWHM values from both this work and Nardetto's are shown in Table 1 for the 6 Cepheids. We used a Gaussian least-square fitting method on our ACF profile at each phase to measure the statistical uncertainties. Nardetto et al. (2006) used a similar measurement process revolving around a bi-gaussian least-square fit. This difference arises from the fact that the ACF is symmetric whereas a single-line profile is not. FWHM statistical uncertainties of Nardetto et al. (2006) were directly taken from their paper and normalized to establish a direct comparison with the ACF technique. Statistical analysis shows that 4 out of the 6 Cepheids (R Tra, RZ Vel, S Cru, Y Sgr) have significantly lower uncertainties on average by using the ACF over a single-line measurement method. On the other hand, RS Pup and Y Oph show closer results between the two techniques, slightly favouring the single-line method at some phases. This can be explained by the fact that at all phases, Y Oph had ACF profiles characterized by strong wings. For this particular Cepheid, a Lorentzian or Voigt function would probably have been the better choice and decreased significantly the uncertainties values.

However, although quantitative, this comparison only gives an approximate estimate of the advantage of the autocorrelation. In principle, the advantage should be determined by comparing the noise in the curves in Fig. 1 obtained with the autocorrelation and by Nardetto et al. (2006). This could be done by fitting theoretical curves to the observational curves in Fig. 1. Unfortunately, this cannot be done, because we do not know what theoretical curves should be used. Note also that the curves in Fig. 1 vary substantially among the Cepheids. However, a qualitative estimate of the advantage of the autocorrelation can be obtained by looking, in Fig. 1, at the phases at which there are at least 2 observations at almost the same phase, We can then see that the values obtained with the autocorrelation are significantly more similar than those obtained by Nardetto et al. (2006). This can be seen, for example, for R Tra and S Cru. Even for the 2 Cepheids, where there are nearly identical results in Table 1 for the 2 techniques, Fig. 1 shows that the autocorrelation gives better results. This can be seen near phase 0.4 for Y Oph and RS Pup.

### 3.2    $\sigma_{obs}$ and $\sigma_{res}$ curves

Results for X Cyg were published by Bersier & Burki (1996) who used the cross-correlation function with a mask. Their spectra were acquired with CORAVEL. The spectrometer is mounted on the 1.54m Danish telescope at La Silla in Chile (Imbert & Prevot 1981).

We analyze 16 spectra of the same star from the SOPHIE (R ∼ 75000 in high resolution mode, 40000 in high efficiency mode) online database using the autocorrelation technique. No previous work has been published for these spectra. SOPHIE is installed on the 1.93m reflector telescope at the Haute-Provence Observatory (Hebrard & The SOPHIE Team 2011). We compare our $\sigma_{obs}$ curve to the one of Bersier & Burki (1996) (Fig. 2). The X Cyg curve was calculated by using a pulsation period of 16.3857 days. The $\sigma_{obs}$ values are calculated from FWHM values with the known Gaussian relation: $FWHM_{obs} = 2.3548\sigma_{obs}$. The autocorrelation of a Gaussian signal remains Gaussian with a FWHM increased by a factor $\sqrt{2}$. We thus divided every $\sigma_{obs}$ value given by the ACF by $\sqrt{2}$ to be consistent with $\sigma_{puls}$ values that were calculated by taking into account this $\sqrt{2}$ multiplicative factor.

Our results are displayed on the left side of the figure while those of Bersier & Burki (1996) are on the right side. Even though we have fewer spectra at our disposal, we observe a smooth curve using the autocorrelation technique. The pattern of the curve resembles the fit made by Bersier & Burki (1996) on X Cyg. The fit was superimposed on the results obtained using the ACF. We measure a slightly more pronounced bump at phase φ ∼ 0.9. If we compare both telescopes (1.54m Danish telescope and 1.93m reflector telescope), we note that they have a similar mirror size. This further demonstrates the utility of the ACF over the CCF. The ACF is an easier tool to use than the CCF and provides great precision.

Out of the 16 spectra, 13 correspond to different pulsational phases. There is a superposition of $\sigma_{obs}$ values from two spectra at phases φ = 0.11, 0.31 and 0.81. These phases are identified in Fig. 2 by a square. The superpositions are so good that one cannot see them in Fig. 2. This superposition confirms the efficiency of the autocorrelation function against photon noise. These results demonstrate that the autocorrelation technique gives very accurate linewidth curves of Cepheids.

We plot in Fig. 3 the $\sigma_{obs}$ and $\sigma_{puls}$ values as a function of phase for the 6 other Cepheids obtained with the HARPS spectrograph. The pulsation periods used to calculate the phase of the spectra are the same as those of Nardetto et al. (2006). The limited phase coverage for some Cepheids (e.g. Y Oph) prevents us from tracing a smooth curve.

Microturbulence curves as a function of pulsational phase have been measured in previous works by several authors using a variety of different techniques (Bersier & Burki 1996; Benz & Mayor 1982; Gillet et al. 1999). Microturbulence measurements may vary depending on the analyzed line because different lines may have different widths. Like the CCF method, the ACF provides an average microturbulence amplitude based on multiple lines. As discussed in section 2, strong lines (e.g. Balmer lines) were removed so that there is no significant problem caused by the average of all the lines. For each Cepheid, we compute $\sigma_{res}$ using Eq. (3). These curves are plotted in Fig. 4. All Cepheids with the exception of Y Oph, due to a lack of data, have a similar $\sigma_{res}$ pattern. We note a sharp peak between phases φ = 0.8-0.9 for most stars. This phase interval corresponds to the moment when the star reaches its maximum microturbulence velocity near its minimum radius. Those results are in agreement with prior works conducted on other Cepheids (Bersier & Burki 1996; Gillet et al. 1999).

### 3.3    Degradation of spectra

To further assess the capabilities of the ACF, we artificially degraded the quality of the spectra. The spectral degradation was done with MATLAB in 3 steps:

- The spectra are first convolved with a Gaussian function. The FWHM of the Gaussian function determines the extent of the simulated instrumental broadening. In this paper, we chose FWHM values so that the resolving power of the spectrograph is reduced by a factor of 5, 10 and 20 (R ∼ 23000, 12000 and 6000 respectively).

- The convolved spectra are then interpolated into fewer pixels to reduce the data sampling. The original HARPS spectra with $R \sim 120000$ (Nardetto et al. 2006) have, on average, 312000 pixels of data sampling. A spectral degradation of a factor 5 would, for example, bring the sampling of the spectra to 62400 pixels.

- A considerable amount of random noise is added to the spectra to simulate photon noise.

Photon noise was simulated with MATLAB software using the function randn which generates values at random between 0 and 1. We first measure the standard deviation of the noise ($\sigma_{noise}$) on each spectra between 5672 and 5674 Å where no line is present. The amplitude of the noise to be added with the randn function is weighted by a multiplier coefficient. This coefficient is set so that we obtain a $\sigma_{noise}$ that is 2.5 times more than the original value.

We show in Fig. 5 FWHM curves as a function of phase for R Tra after degradation. As the severity of the spectral degradation increases, the ratio between the maximum and minimum values decreases. We shall improve the results by removing in the degraded spectra the instrumental broadening, from the autocorrelation, by Gaussian approximation.

Results are greatly improved after removing the instrumental broadening term from Eq. (3). The amplitude of variations are nearly completely restored (see Fig. 6). A degradation by a factor of 5 does not significantly change the variation curve when compared to the undegraded curve. When degraded by a factor of 20, the resulting curve is slightly less precise than the original but the periodic variation still remains easily detectable. Let us recall that at this level of degradation, the spectra have an equivalent resolving power of only 6000. Moreover, a large amount of random noise was added to the spectra to further disrupt the quality of the spectra. Fig. 6 shows how powerful the autocorrelation technique can be.

## 4    DISCUSSION AND CONCLUSION

In this paper, we presented an autocorrelation technique that we applied to the spectra of Cepheids in order to extract precise microturbulence curves. We compared our variation curves of the FWHM as a function of phase with those of Nardetto et al. (2006) who used a single-line measurement method. We note an improvement in the shape of the curves of all Cepheids in this work using the autocorrelation function.

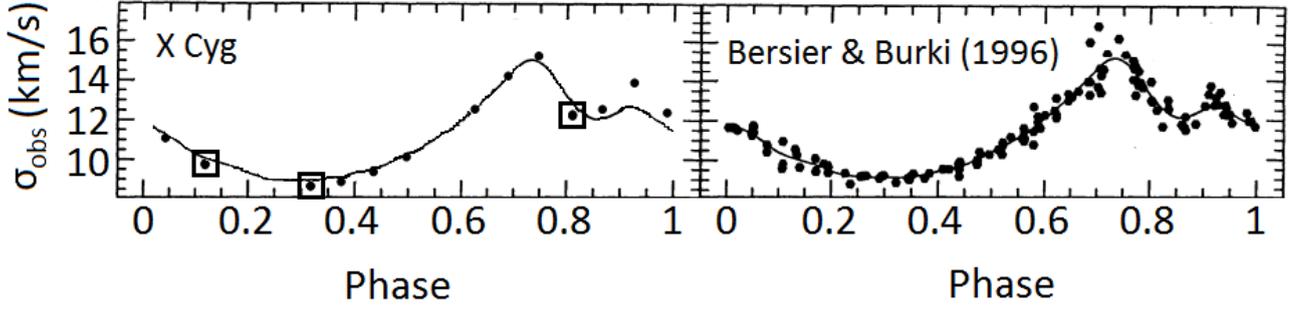

**Figure 2.** The case of the Cepheid X Cyg. Comparison between the $\sigma_{obs}$ curve obtained using the autocorrelation function (left side) and the one obtained by Bersier & Burki (1996) using the cross-correlation technique (right side). In our figure, there are superpositions of $\sigma_{obs}$ values from two spectra at phases $\varphi$ = 0.11, 0.31 and 0.81 (squares).

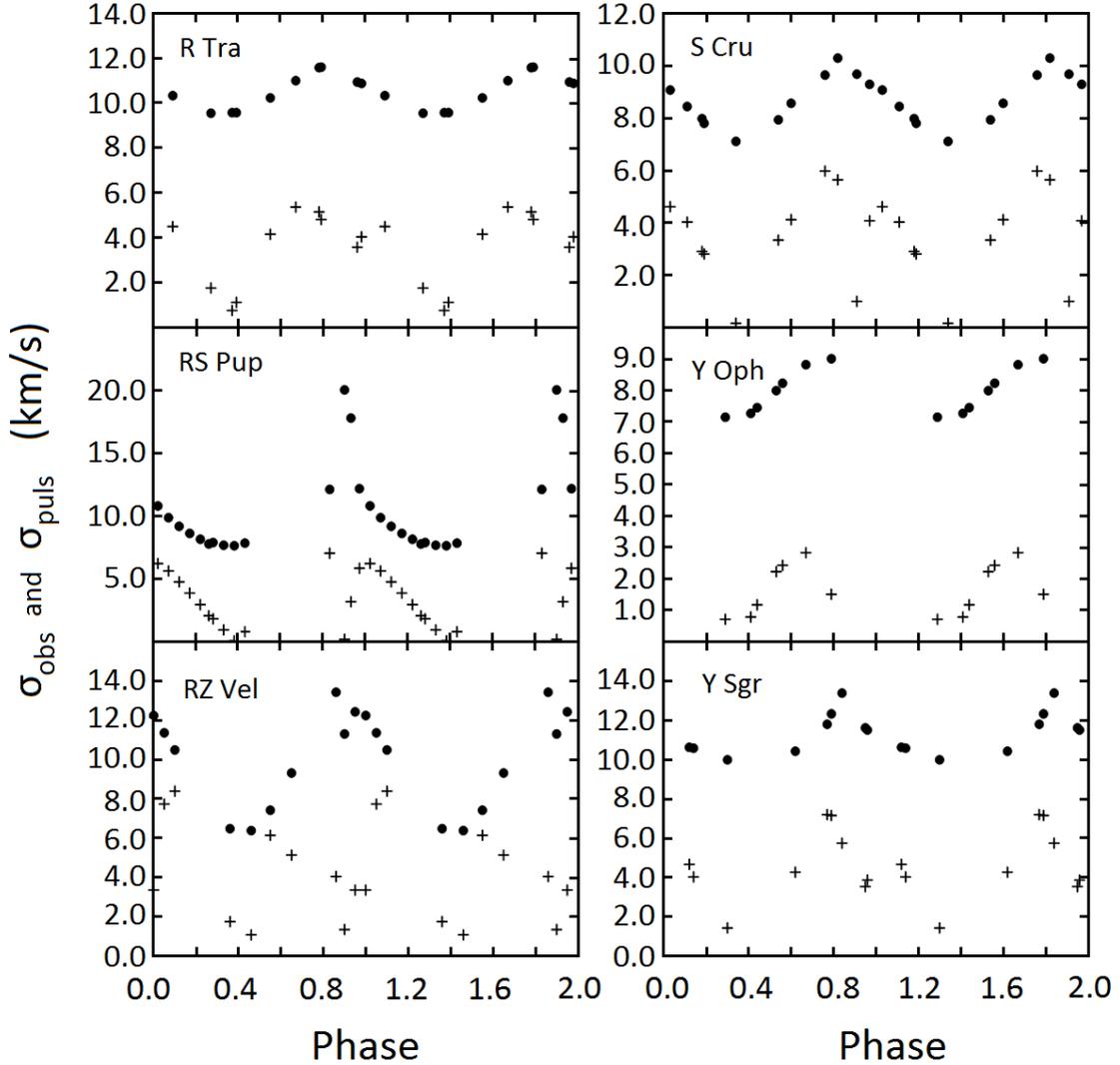

**Figure 3.** $\sigma_{obs}$ (dots, upper values) and $\sigma_{puls}$ (crosses, lower values) as a function of phase for 6 Cepheids.

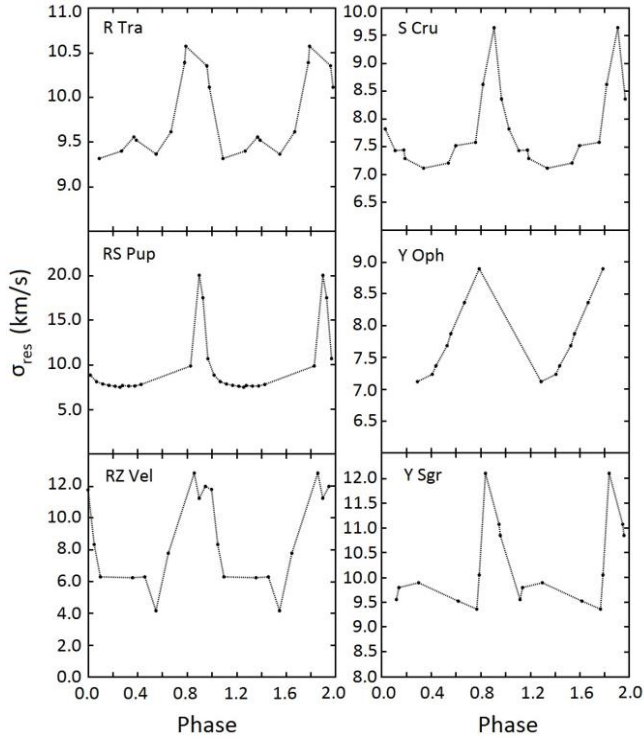

**Figure 4.** σ$_{res}$ computed for 6 Cepheids using Eq. (3).

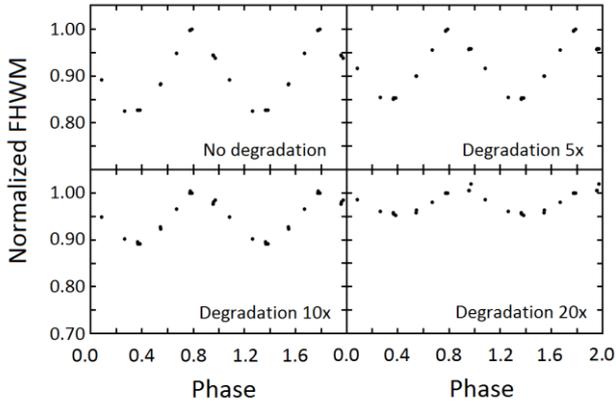

**Figure 5.** Cepheid R Tra. FWHM curve of original spectra with no degradation, degraded 5x, degraded 10x and degraded 20x.

For all Cepheids, with the exception of Y Oph, we obtained a sharp peak in the microturbulence curves around phase φ = 0.9. The peak coincides with the moment when the Cepheid is near its minimum radius. This confirms that in this phase of pulsation, the expansion mechanism attributed to a microturbulence velocity phenomenon is dominant. The case of Y Oph was more difficult to analyze due to the lack of available data.

The comparison of our results obtained for X Cyg with the ACF to results obtained with the CCF, discussed in section 3.2, show that the ACF gives better results than the CCF. The spectra of Cepheids were artificially degraded to further assess the extent of the advantages of the ACF (see section 3.3). After removal of the instrumental broadening, the degradations by a factor of 5 and 10 did not change the results. A degradation by a factor of 20 slightly affected the results but an average linewidth variation remains easily detectable. These results demonstrate the efficiency of the autocorrelation function, especially when applied to spectra of low spectral resolution.

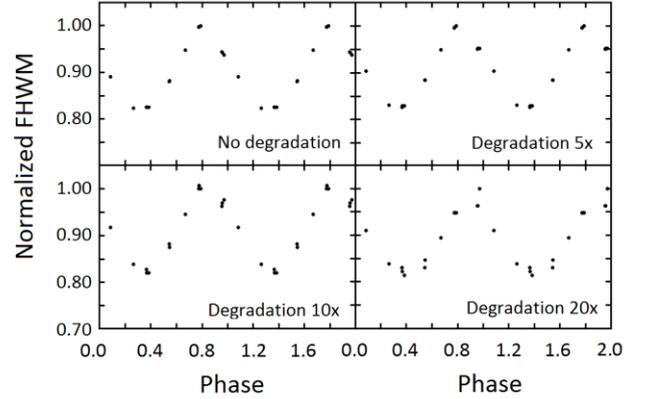

**Figure 6.** Cepheid R Tra with removal of instrumental broadening. FWHM curve of original spectra with no degradation, degraded 5x, degraded 10x and degraded 20x.

The autocorrelation technique could become a worthy tool in many astrophysical fields. It is straightforward, easy to use and offers excellent performance. Further work will be done to investigate this new technique.

**ACKNOWLEDGEMENTS**

This research was supported by the Natural Sciences and Engineering Research Council of Canada.

.

**Table 1.** σ<sub>obs</sub>, normalized FWHM and normalized FWHM (Nardetto et al. 2006) values with statistical uncertainties as a function of phase for 6 Cepheids.

| Phase ($\varphi$) | $\sigma_{obs}$ (km s$^{-1}$) | Norm. FWHM | Norm. FWHM (Nardetto et al. 2006) |
|---|---|---|---|
| | | R Tra | |
| 0.09 | 10.35 ± 0.06 | 0.892 ± 0.005 | 0.875 ± 0.017 |
| 0.27 | 9.56 ± 0.06 | 0.825 ± 0.005 | 0.846 ± 0.009 |
| 0.37 | 9.59 ± 0.07 | 0.826 ± 0.006 | 0.841 ± 0.004 |
| 0.39 | 9.57 ± 0.07 | 0.824 ± 0.006 | 0.837 ± 0.006 |
| 0.55 | 10.25 ± 0.06 | 0.881 ± 0.005 | 0.861 ± 0.009 |
| 0.67 | 11.08 ± 0.07 | 0.948 ± 0.006 | 0.921 ± 0.015 |
| 0.78 | 11.60 ± 0.07 | 0.998 ± 0.006 | 1.000 ± 0.009 |
| 0.79 | 11.62 ± 0.07 | 1.000 ± 0.006 | 0.981 ± 0.013 |
| 0.96 | 10.96 ± 0.05 | 0.945 ± 0.004 | 0.901 ± 0.007 |
| 0.98 | 10.89 ± 0.05 | 0.938 ± 0.004 | 0.942 ± 0.013 |
| | | RS Pup | |
| 0.02 | 10.82 ± 0.04 | 0.539 ± 0.002 | 0.453 ± 0.003 |
| 0.07 | 9.88 ± 0.04 | 0.492 ± 0.002 | 0.414 ± 0.002 |
| 0.12 | 9.19 ± 0.04 | 0.458 ± 0.002 | 0.387 ± 0.001 |
| 0.17 | 8.62 ± 0.06 | 0.430 ± 0.003 | 0.359 ± 0.001 |
| 0.22 | 8.16 ± 0.08 | 0.406 ± 0.004 | 0.321 ± 0.001 |
| 0.26 | 7.78 ± 0.08 | 0.388 ± 0.004 | 0.310 ± 0.001 |
| 0.28 | 7.90 ± 0.08 | 0.393 ± 0.004 | 0.332 ± 0.001 |
| 0.33 | 7.68 ± 0.08 | 0.383 ± 0.004 | 0.282 ± 0.001 |
| 0.38 | 7.64 ± 0.08 | 0.381 ± 0.004 | 0.284 ± 0.001 |
| 0.43 | 7.85 ± 0.10 | 0.391 ± 0.005 | 0.276 ± 0.001 |
| 0.83 | 12.13 ± 0.04 | 0.604 ± 0.002 | 0.523 ± 0.004 |
| 0.90 | 20.07 ± 0.06 | 1.000 ± 0.003 | 1.000 ± 0.010 |
| 0.93 | 17.82 ± 0.12 | 0.888 ± 0.006 | 0.713 ± 0.012 |
| 0.97 | 12.19 ± 0.06 | 0.607 ± 0.003 | 0.502 ± 0.005 |
| | | RZ Vel | |
| 0.00 | 12.26 ± 0.05 | 0.912 ± 0.004 | 1.000 ± 0.020 |
| 0.05 | 11.38 ± 0.07 | 0.846 ± 0.005 | 0.777 ± 0.015 |
| 0.10 | 10.50 ± 0.07 | 0.781 ± 0.005 | 0.781 ± 0.030 |
| 0.36 | 6.48 ± 0.05 | 0.482 ± 0.004 | 0.372 ± 0.002 |
| 0.46 | 6.39 ± 0.08 | 0.475 ± 0.006 | 0.393 ± 0.002 |
| 0.55 | 7.43 ± 0.07 | 0.552 ± 0.005 | 0.628 ± 0.003 |
| 0.65 | 9.33 ± 0.07 | 0.694 ± 0.005 | 0.895 ± 0.005 |
| 0.86 | 13.45 ± 0.04 | 1.000 ± 0.003 | 0.699 ± 0.003 |
| 0.90 | 11.32 ± 0.05 | 0.842 ± 0.004 | 0.821 ± 0.005 |
| 0.95 | 12.45 ± 0.05 | 0.926 ± 0.004 | 1.000 ± 0.015 |

**Table 1** – *continued* $\sigma_{obs}$, normalized FWHM and normalized FWHM (Nardetto et al. 2006) values with statistical uncertainties as a function of phase for 6 Cepheids.

| Phase ($\varphi$) | $\sigma_{obs}$ (km s$^{-1}$) | Norm. FWHM | Norm. FWHM (Nardetto et al. 2006) |
|---|---|---|---|
| | | S Cru | |
| 0.03 | 9.08 ± 0.05 | 0.881 ± 0.005 | 0.823 ± 0.010 |
| 0.11 | 8.46 ± 0.04 | 0.820 ± 0.004 | 0.800 ± 0.010 |
| 0.18 | 7.99 ± 0.05 | 0.776 ± 0.005 | 0.733 ± 0.004 |
| 0.19 | 7.81 ± 0.05 | 0.758 ± 0.005 | 0.599 ± 0.009 |
| 0.34 | 7.12 ± 0.04 | 0.691 ± 0.004 | 0.652 ± 0.002 |
| 0.54 | 7.94 ± 0.05 | 0.771 ± 0.005 | 0.748 ± 0.004 |
| 0.60 | 8.58 ± 0.05 | 0.832 ± 0.005 | 0.823 ± 0.004 |
| 0.76 | 9.66 ± 0.04 | 0.937 ± 0.004 | 0.968 ± 0.010 |
| 0.82 | 10.31 ± 0.04 | 1.000 ± 0.004 | 1.000 ± 0.009 |
| 0.91 | 9.69 ± 0.05 | 0.940 ± 0.005 | 0.878 ± 0.017 |
| 0.97 | 9.30 ± 0.06 | 0.903 ± 0.006 | 0.855 ± 0.009 |
| | | Y Oph | |
| 0.29 | 7.16 ± 0.06 | 0.793 ± 0.007 | 0.779 ± 0.004 |
| 0.41 | 7.28 ± 0.06 | 0.807 ± 0.007 | 0.768 ± 0.008 |
| 0.44 | 7.46 ± 0.06 | 0.827 ± 0.007 | 0.848 ± 0.004 |
| 0.53 | 8.00 ± 0.07 | 0.887 ± 0.008 | 0.905 ± 0.004 |
| 0.56 | 8.24 ± 0.06 | 0.913 ± 0.007 | 0.928 ± 0.004 |
| 0.67 | 8.83 ± 0.07 | 0.979 ± 0.008 | 1.000 ± 0.004 |
| 0.79 | 9.02 ± 0.07 | 1.000 ± 0.008 | 0.985 ± 0.008 |
| | | Y Sgr | |
| 0.12 | 10.63 ± 0.05 | 0.793 ± 0.004 | 0.775 ± 0.018 |
| 0.14 | 10.59 ± 0.05 | 0.790 ± 0.004 | 0.780 ± 0.006 |
| 0.30 | 10.00 ± 0.03 | 0.746 ± 0.002 | 0.754 ± 0.005 |
| 0.62 | 10.44 ± 0.04 | 0.779 ± 0.003 | 0.746 ± 0.008 |
| 0.77 | 11.81 ± 0.07 | 0.881 ± 0.005 | 0.903 ± 0.011 |
| 0.79 | 12.34 ± 0.08 | 0.921 ± 0.006 | 0.928 ± 0.010 |
| 0.84 | 13.40 ± 0.04 | 1.000 ± 0.003 | 1.000 ± 0.011 |
| 0.95 | 11.63 ± 0.04 | 0.868 ± 0.003 | 0.847 ± 0.005 |
| 0.96 | 11.52 ± 0.04 | 0.860 ± 0.003 | 0.826 ± 0.006 |